# Modelling Quantum Circuits with UML

Ricardo Pérez-Castillo
University of Castilla-La Mancha
Talavera de la Reina, Spain
ricardo.pdelcastillo@uclm.es

Luis Jiménez-Navajas
University of Castilla-La Mancha
Ciudad Real, Spain
luis.jimeneznavajas@uclm.es

Mario Piattini
University of Castilla-La Mancha
Ciudad Real, Spain
mario.piattini@uclm.es

## ABSTRACT

None of the quantum computing applications imagined will ever become a reality without quantum software. Quantum programmes have, to date, been coded with ad hoc techniques. Researchers in the field of quantum software engineering are, therefore, now demanding more systematic techniques and methods with which to produce software with sufficient quality. One of the challenges and lessons learned from classic software engineering is the need for high-level, abstract and technology-independent representations with which to design software before it is coded. This paper specifically addresses this challenge for quantum software design. Since UML is a well-proven modelling language that has been widely employed by industry for some time, we propose a UML extension for the representation of quantum algorithms. Our proposal comprises the definition of a UML profile based on various stereotypes that can be applied to the existing UML activity diagrams in order to represent quantum circuits. The advantage of this representation is that UML quantum circuits can be interrelated with other UML elements and diagrams, which will make it possible to represent various concerns and viewpoints of the so-called hybrid information systems. This will consequently enable classical and quantum aspects to be modelled together in integrated designs in a technological-agnostic manner that is already supported by a considerable number of existing software design tools.

## 1 INTRODUCTION

The impact of quantum computing on today's society is undeniable [1, 2]. Quantum computing has a many promising applications [3], such as cryptography, financial services, pharmacy and health, energy and farming, defence, etc.

None of the advantages that have been forecast to appear with the advent of quantum computing will be achieved with cutting-edge machines only, but these applications could be brought into existence through the use of quantum software [4]. This new computing paradigm has devised a completely different approach for programming, along with building blocks for quantum programmes that are also unique. Quantum software is typically designed as quantum circuits that apply a set of quantum gates to various qubits in order to explore a search space in a non-deterministic and probabilistic manner.

In this scenario, quantum software engineering needs to be developed as a new field [4, 5] in order to provide new techniques, methods and practices with which to analyse, design, code and create quantum software with sufficient quality in a more systematic manner. Other well-proven and successful classical software engineering techniques and methods could, together with these new techniques and methods, be adapted for quantum software [5]. The adaptation of existing techniques and methods is also important owing to the coexistence of classical and quantum software, which in many cases operate together in the so-called hybrid information systems.

We believe that hybrid information systems will become mainstream as quantum computing machines are improved and more and more companies invest in migrating parts of their classical software towards quantum. However, it does not, from an economic point of view, make sense to implement every tiny and simplistic business process as quantum software, since classical software still performs better in the case of certain problems. Our envisioned scenario considers companies that migrate some of their mission-critical functionalities to quantum software while other new functionalities are implemented in quantum software owing to the new possibilities facilitated by this new computing paradigm. Classical and quantum software should, therefore, be modernised in order to attain hybrid information systems [6].

Many of the problems solved by the existing software engineering methods and techniques are still the same as those involved in the design and construction of hybrid information systems [7]. For example, abstract representations for software are a key aspect as regards discussing design concerns and modelling systems with high-level representations, while implementation details are hidden. One well-proven solution, that is most widely used in classical software engineering, is the usage of standard modelling languages such as UML [8] for the analysis and design of information systems. We believe that UML modelling is a powerful tool for the design of hybrid information systems. This can be achieved by following a model-driven engineering (MDE) approach that has at least two important advantages. First, UML models focus on domain and conceptual representations in a technological-agnostic manner, and second, automated model transformations can be established from/to UML to/from source code for different platforms. Quantum software engineering that designs and develops quantum software by means of UML can consequently abstract technical complexities while focusing on the domain of the problem or business model, thus requiring only the functional knowledge needed for the solution. The advantages of MDE are a key aspect at this time of rapid evolution and a lack of standardisation in quantum programming, since companies are afraid of investing in platforms that will not continue in the future.

Although the UML was defined in a general and technology-independent manner, it was not originally conceived for the design of quantum software. It is, therefore, necessary to extend UML in order to cover the new quantum concerns. This paper introduces the ongoing results of research focused on the extension of the UML and presents a preliminary UML profile with which to represent quantum circuits as activity diagrams. The main implication of this work is that quantum circuits (based on the variant of the Penrose graphical notation) can be represented with UML, signifying that these elements can be linked with other abstract design elements of hybrid information systems that are also represented in the UML through the use of existing [9] or future extensions.

The remainder of the paper is structured as follows: Section 2 states the relevance of using UML in quantum software engineering, after which Section 3 introduces the UML profile for use in modelling quantum circuits, with a running example for the quantum teleportation algorithm. Finally, Section 4 discusses the main implication and future efforts of this research.

## 2 UML FOR QUANTUM SOFTWARE

This section presents the usage of UML in quantum software engineering and how it can be extended.

### 2.1 Usage of UML in Quantum Software

The development of quantum or hybrid information systems cannot simply consist of a collection of code modules. The development of these systems should rather follow a whole life cycle, i.e. a "pre-defined pathway for implementing and solving large projects on quantum both in a time-efficient and resource-efficient manner" [7]. It does not matter how long the life cycle is, since it is certain that the quantum software must be designed at some point. Software design defines the architecture, system elements, interfaces and other characteristics of a system [10] in order to accomplish goals using a set of primitive components, and is subject to constraints [11].

UML can help by gathering and analysing software requirements and incorporating them into a programme design in a technology- and methodology-independent manner. This will make it possible to additionally use UML with hybrid information systems.

Although other modelling languages can be used to design software, we believe that the usage of UML in quantum software engineering will have several advantages:

1. **Different perspectives**. UML provides many different diagrams types to look at systems from various perspectives and represent different concerns. These viewpoints are useful as regards modelling hybrid information systems.
2. **Design validation.** The aforementioned perspectives allow UML to help quantum software engineers to communicate, explore potential designs and validate the architectural design of the software. The UML is highly extended in industry and is, in some cases and to a certain extent, easy for non-technical staff to understand.
3. **Best practices.** UML represents a collection of best engineering practices that have proved successful in the modelling of large complex systems. These practices could consequently be applied in quantum/hybrid information systems. One example of this is the aforementioned MDE approach, which ensures platform independence.
4. **Structured Design.** UML modelling makes it easier to structure software as a collection of self-contained modules or components. This enables the reuse of code, scalability, and robustness. The state of the art of the quantum software engineering field is demanding precisely this [5].
5. **Tooling**. Since UML is a widely adopted ISO/IEC standard, most of the design and modelling tools support it. One of the primary goals of the UML is to advance the state of industry by enabling object visual modelling tool interoperability [8]. Quantum software modelling could be integrated into the tools used by many software engineers.
6. **Software Modernisation.** UML is not only used for designing target hybrid systems that will then be implemented by forward engineering. UML models can also be generated by reverse engineering tools that analyse existing software, e.g., in order to migrate or modernise software towards hybrid information systems [6].

Despite these advantages, UML needs to be adapted in order to capture all the new semantics and building blocks involved in quantum software. Literature already contains some first approximations. For example, in [9], Q-UML is proposed as a concrete syntax definition with which to represent certain quantum elements in class and sequence diagrams. In [6], UML is stated to be a relevant model for use in software modernisation processes, such as the reverse engineering or restructuring phase, and a UML extension is introduced with an example for use case diagrams. Other authors have already used UML (without providing extensions) to model quantum software [12].

### 2.2 UML Extensibility

UML was defined on the basis of the MOF (Meta-Object Facility), which is a meta-metamodel. UML is, therefore, a metamodel that is used to define different UML models, and the extension of UML consequently consists of extending the metamodel. It is necessary to bear in mind that all metamodels have both an abstract syntax (that describes the concepts in the language, their characteristics and interrelationships) and a concrete syntax (that defines the specific textual or graphical notations required for the abstract elements). It is possible to extend the UML by principally following three different approaches [13].

1. **A new instance of the MOF model**. This approach consists of creating a completely new metamodel based on MOF. The result of this heavyweight approach is a new Domain-Specific Modelling Language (DSML).
2. **Derivation of a new UML metamodel**. This approach adds new metamodel elements to the existing one. As occurs with the first approach, it is a different metamodel, but at least considers the original UML metamodel as it is.
3. **UML Profile**. This is a lightweight extension approach that is based on the UML built-in extension mechanism, UML Profiling. UML profiles are created as a set of stereotypes, tagged values and constraints defined for some of the existing UML elements.

These three approaches have various pros and cons. The expressiveness of the two first approaches is powerful, since conformity with UML is not necessary (particularly in the case of approach 1). Despite the fact that the expressiveness of UML profiles is limited, standardisation and conformance are better, since the extension is fully compliant with UML. This advantage is a key aspect as regards the usage of the defined profile with existing UML modelling tools. Moreover, it is easier to maintain extensions that have been defined as UML profiles since the associated modelling tools do not need to be adjusted after each change, as occurs with a DSML. DSMLs (approaches 1 and 2), in fact, usually end up with an overloaded and imprecise language. The aforementioned advantages lead us to believe that the UML profile is the best way in which to define the UML extension for quantum information systems.

Figure 1 presents the part of the UML metamodel (abstract syntax) employed to define UML Profiles. A UML profile is defined as a package containing a set of defined *stereotypes* (that may or may not have a specific image). The UML profile must then be applied to a certain model. The attributes used to filter which UML elements are available when the Profile is applied are *metamodelReference* and *metaclassReference*. When a stereotype is applied to a model element, the values of the properties are traditionally referred to as *tagged values*.

## 3 QUANTUM UML PROFILE

This section presents the preliminary UML profile defined in order to represent quantum programmes as activity diagrams by using a graphical notation similar to that employed by quantum circuits. It should be noted that a broader UML profile will be defined for the representation of not only quantum circuits but also the analysis and design concerns of hybrid information systems. These other UML diagrams are not, however, within the scope of this paper.

The entire Quantum UML Profile is presented in Figure 2. The UML profile consists of 6 stereotypes with which to add the new semantic related to quantum circuits (see dark gray elements at right-hand side of Figure 2): *quantum circuit, qubit, quantum gate, control qubit, measure, reset.* The left-hand side of Figure 2 shows an excerpt of the UML metamodel employed to represent UML Activity Diagrams, i.e., the base diagram used to model quantum circuits with UML.

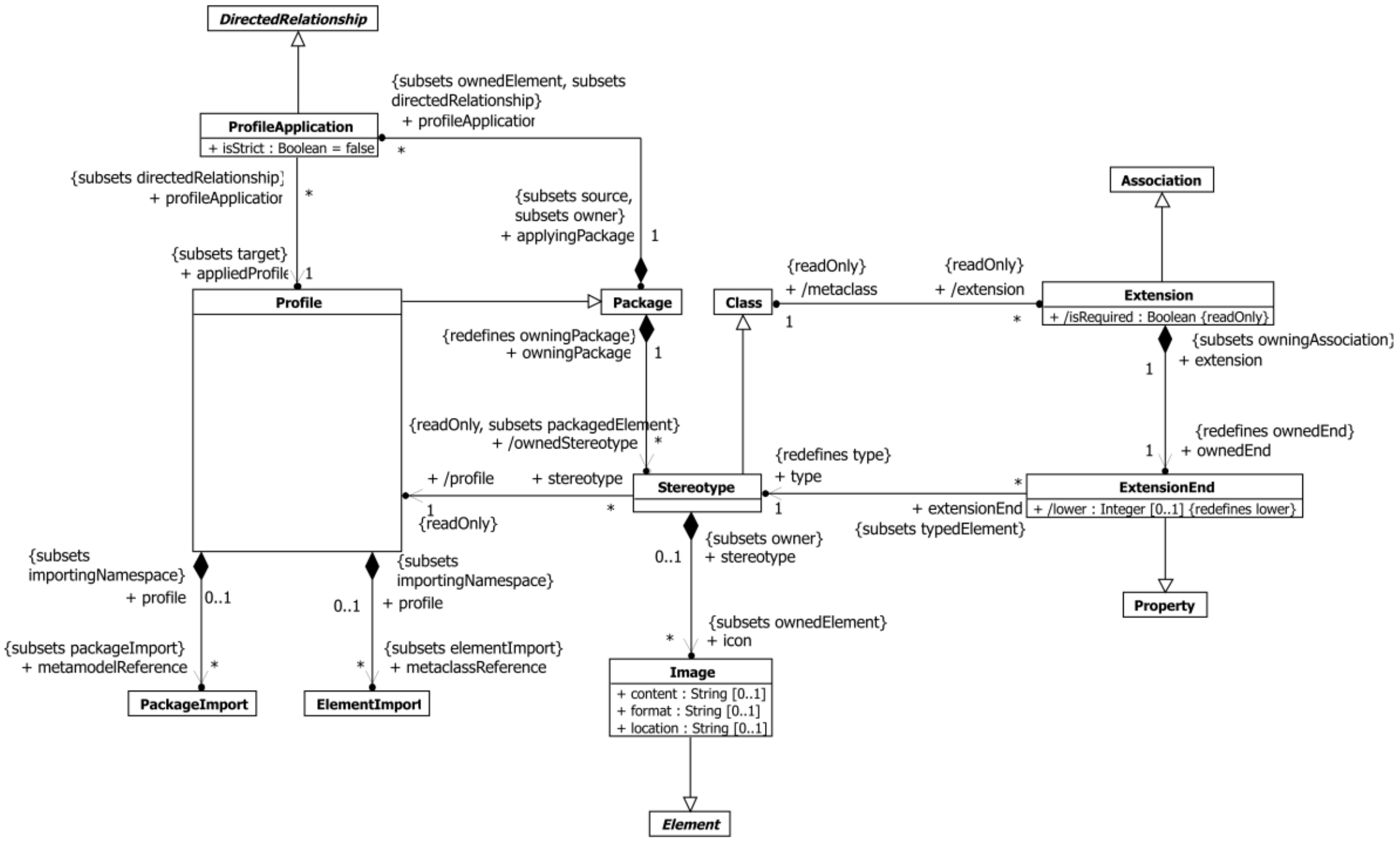

**Figure 1:** UML metamodel employed to define UML Profiles (Adapted from [8]).

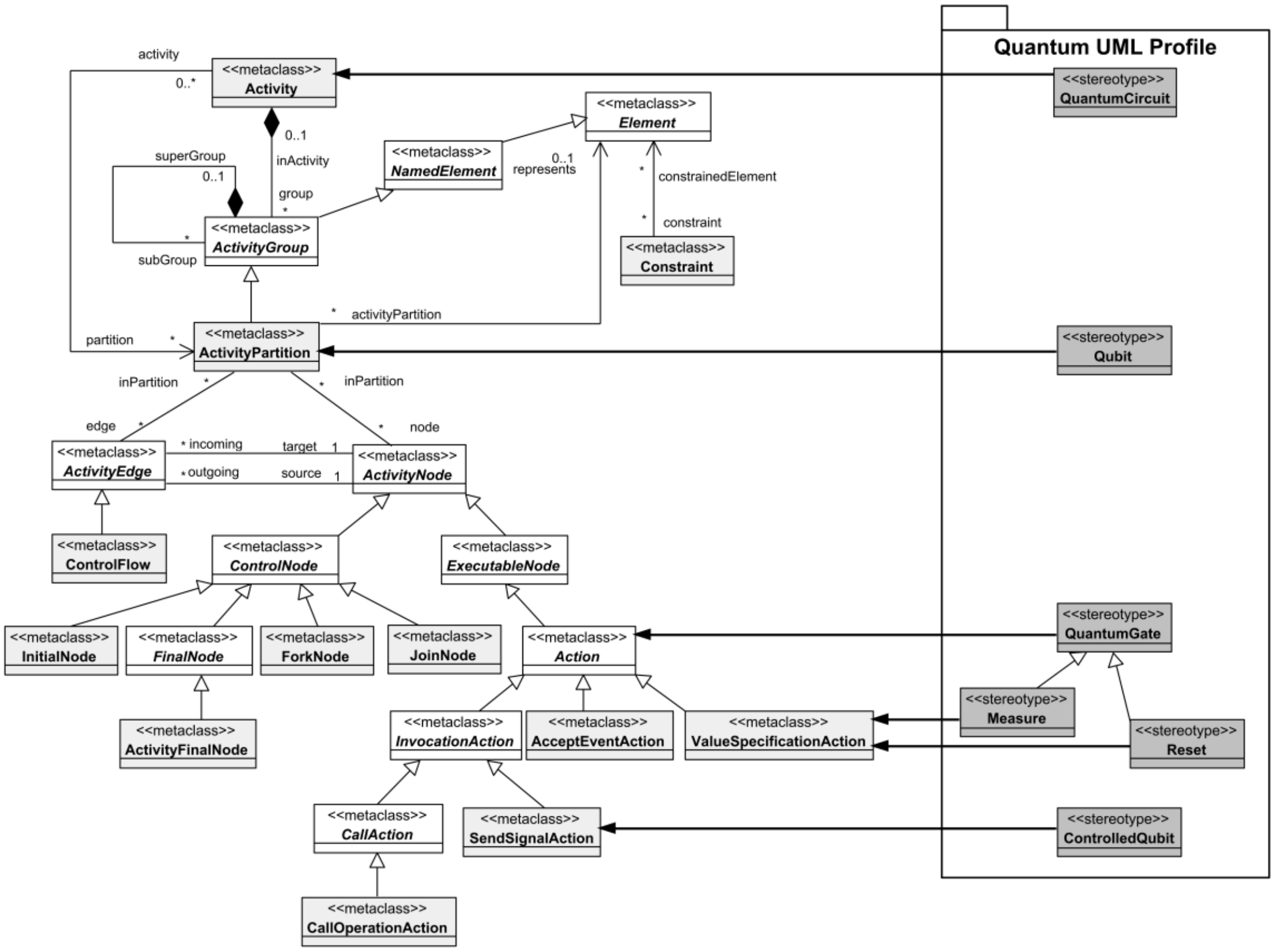

**Figure 2: Quantum UML Profile.**

The metaclass elements (light gray) in Figure 2 are those that are not abstract and are, therefore, the elements available to be included in the UML Activity Diagram. The leftwards arrows from stereotypes to metaclass elements in Figure 2 are *extension* elements (see Figure 1) that are used to indicate that the properties of a metaclass are extended through the use of the respective stereotype.

The intuitive idea behind representing quantum circuits with UML activity diagrams is that each algorithm is represented with a single compound *activity* with the stereotype *<<quantum circuit>>*. The entire circuit is, therefore, defined in this activity, and the compound activity can be reused in other circuits, as occurs in quantum programming. The various *activity partitions* (graphically represented as horizontal swim lanes) can be defined in the parent activity by employing the *<<qubit>>* stereotype. The circuit has as many *activity partitions* as different qubits used in the algorithm. All the different quantum gates applied in the circuits are, therefore, represented as *action* elements and are placed in the respective swim lane, according to the qubit under which the gate is applied or controlled. On the one hand, ordinary quantum gates (such as H, Y, Z, etc.) are represented as *call operation actions* plus the *<<quantum gate>>* stereotype. On the other hand, conditional gates are represented with multiple action elements. The control qubits are represented with *send signal action* elements with the stereotype *<<controlled qubit>>*, while the gate applied is represented with the counterpart element, *accept event action*, plus the *<<quantum gate>>* stereotype (see Figure 2). Additionally, in order to add the semantic concerning the relationships between the control qubits, various UML *constraint* elements are established between the *action* elements involved.

In addition to these core elements, special operations, such as qubit measuring and qubit resetting, are represented with *value specification action* elements and their respective stereotypes *<<measure>>* and *<<reset>>*.

The control flow of quantum circuits is represented in the UML Activity Diagram with *control flow* elements that connect two *action* elements. In quantum circuits, isolated quantum gates that are applied independently in different qubits can sometimes be executed in any order. In this case, a control flow is established from top to bottom for every qubit. On other occasions, the order of certain quantum gates is important, and barriers are used in graphical quantum circuits. In this case, these synchronizations are represented in the UML Activity Diagram with *fork* and *join nodes.* The control flow in UML Activity Diagrams should eventually be defined in a continuous manner, starting from the special element *initial node*, and ending in the special element *activity final node.* This signifies that the result should be a fully connected graph. This is a change as regards graphical quantum circuits, in which the control flow could, to a certain extent, be ambiguous. The UML extensions we provide support and advocate the definition of an exact control flow, similar to that provided by quantum programming languages such as Q# or QASM.

## 3.1 Running Example

In order to illustrate how the Quantum UML Profile is applied, and to demonstrate its applicability, the paper provides a running example by using the teleportation algorithm [14] (see Figure 3).

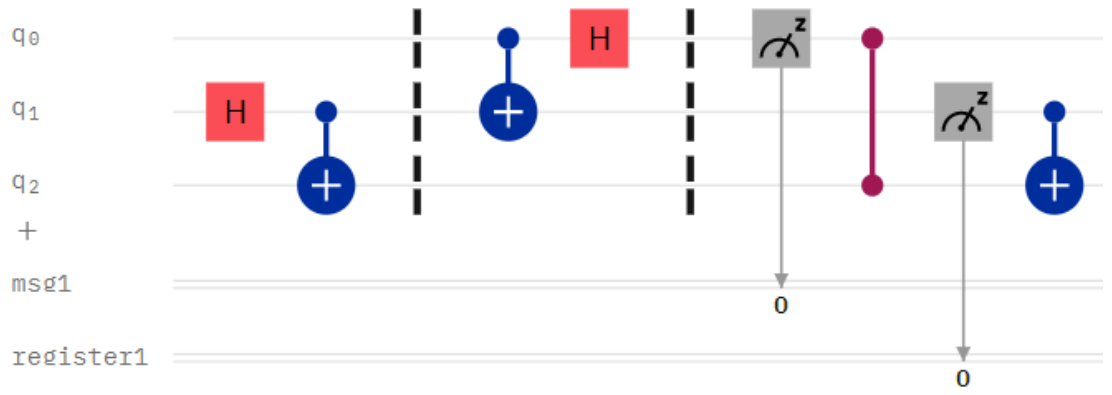

**Figure 3: Quantum circuit for the teleportation algorithm.**

This algorithm supports quantum teleportation, a technique used to transfer quantum information from source to destination by employing entangled states. In this example, q0 is the qubit that represents the message to be sent, q1 is an auxiliary qubit, and q2 is the target qubit that will receive the information coded in q0. In this example, Hadamard (H) gates are used in order to place a qubit into a state of superposition. CNOT is a conditional X gate that rotates the position of the qubit in the X axis (like a NOT gate for classical computers) if the value of another qubit is one. It similarly uses a conditional Z gate that applies a rotation in the Z axis depending on the value of another qubit. In addition to these gates, two measures are taken in qubit q0 (message) and q1 (auxiliary) that collapse these qubits and take certain values. At the end of the algorithm, q2 (target) will have the same value that q0 had.

Figure 4 shows how the Quantum UML Profile is applied in an activity diagram in order to represent the equivalent quantum circuit for the teleportation algorithm. As explained previously, the whole circuit is enclosed in a composed *activity* with the respective stereotype. This circuit, therefore, has three horizontal *activity partitions* (one for each qubit). The quantum gates and measures are then placed as *action* elements in the same position as in the original quantum circuit (compare Figure 3 and Figure 4). The usage of a *fork* and *join* elements for the original synchronization barriers should also be noted. With regard to the quantum gates *CNOT* and *CZ*, these are modelled with pairs of send signal action and accept event action connected by a *restriction* edge (together with the stereotypes *<<controlled qubit>>* and *<<quantum gate>>*).

One interesting aspect of the quantum circuit represented with UML is that measures can be connected with *data store nodes* that represent classical values after a qubit is measured (see *msg1* and *register1* in Figure 4). Other ordinary UML elements, outside the whole circuit, could be connected with elements of the quantum circuit in order to define relationships with classical elements. This is specifically valuable as regards representing three relevant concerns:

- **Quantum requests** from the classical programmes (also known as *drivers*) to the quantum programmes, i.e., the remote calls from the master server.
- **Cost functions** that manage the multiple calls to the quantum circuits and the aggregation of results in the classical source code of the drivers. This is a key aspect, since the non-deterministic and probabilistic nature of quantum algorithms makes it necessary to execute the quantum circuits multiple times.
- **Optimizers** are other functions that are interesting to model in association with the quantum circuits. These functions are used to invoke quantum circuits with different parameters with the goal of optimizing certain circuits (e.g., reduction of quantum gates or qubits).

Finally, with regard to the running example, the execution flow of the circuit in UML is explicitly represented through the use of *control flow* elements from the *initial node* to the *activity final node* through all the quantum gates (see Figure 4). This explicit flow contrasts with the original quantum circuit in Figure 3, in which the execution flow is implicit. If attention is paid to the equivalent QASM source code of the teleportation algorithm (see Figure 5), it will be noted that the explicit control flow modelled with the Quantum UML Profile is almost the same as that defined using the source code (QASM or any other quantum programming language). The explicit execution flow has some advantages in some cases, such as the optimization of quantum algorithms, during which the specific order of the quantum gates may be of interest.

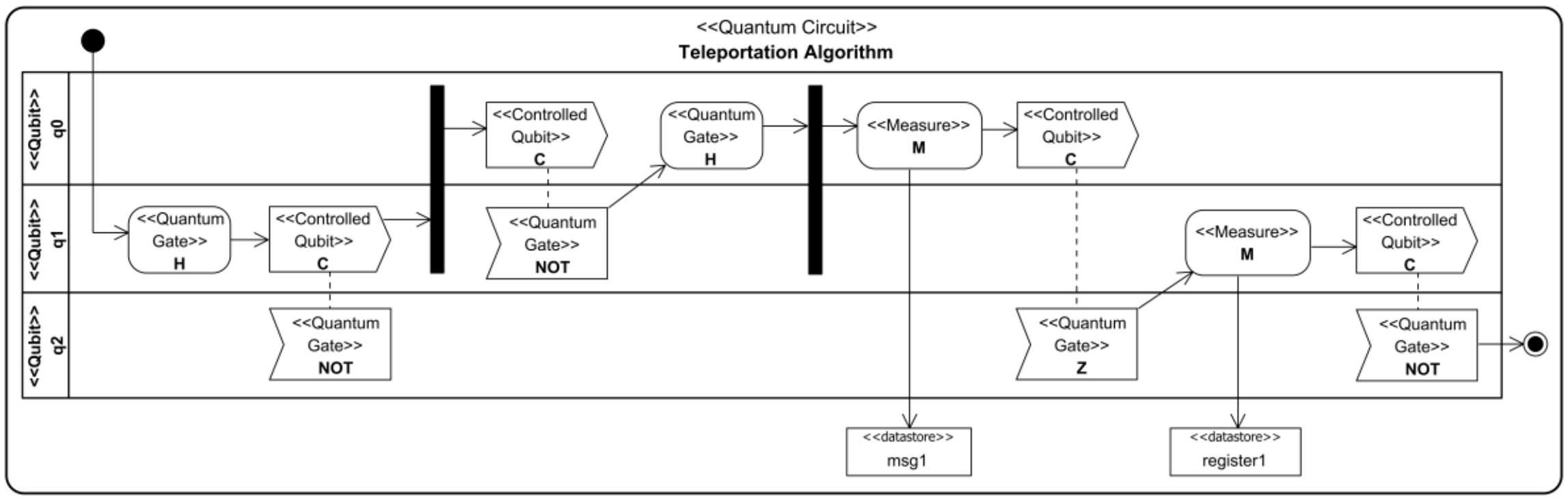

**Figure 4: Quantum circuit represented with UML for the teleportation algorithm.**

```
OPENQASM 2.0;
include "qelib1.inc";
qreg q[3];
creg msg[1];
creg register[1];
h q[1];
cx q[1], q[2];
barrier q[0],q[1],q[2];
cx q[0], q[1];
h q[0];
barrier q[0],q[1],q[2];
measure q[0] -> msg[0];
cz q[0], q[2];
measure q[1] -> register[0];
cx q[1], q[2];
```

**Figure 5: The QASM code for the teleportation circuit.**

# 4 DISCUSSION

This paper introduces the idea of modelling quantum circuits in UML Activity Diagrams. Although several quantum circuit notations with which to graphically represent quantum algorithms already exist, the UML adds a similar notation that is understood by main role players in the quantum software engineering field and which is available in many existing modelling tools. In fact, the approach followed in this research consists of a lightweight extension based on a UML profile (the built-in extension mechanism provided by UML). Unlike other heavy-weight extension mechanisms, we believe the Quantum UML Profile has two clear advantages: (i) the reuse of existing UML modelling tools, and (ii) integration with other standard UML elements, which is useful as regards representing hybrid information systems.

The Quantum UML Profile designed in this paper consists of 6 stereotypes that can be applied to various standard UML elements used in activity diagrams. These stereotypes have been intentionally defined without a graphical icon, as would have been possible. This design decision was made in order to preserve the UML profile in order to make it as aseptic as possible. Literature shows that there is a certain variation in graphical representations of quantum gates. For example, the *CZ* gate is represented with a bullet point for the control qubit connected to a square labelled with 'Z', or can alternatively be represented with two bullet points that are connected (as shown in Figure 3). It was for this reason that we decided to avoid specific graphical representations of the stereotypes defined. Thus, all the quantum gates are supported, and the ordinary graphical UML notation can additionally be associated with similar UML elements (with the same graphical notation) in order to model classical software parts. The modelling of hybrid information systems with a common notation can consequently be improved through a reduction in complexity, thus attaining a better understandability.

The main implication of this work is that quantum circuits can be designed and modelled in the UML. The existing UML-based code generators could, therefore, be extended in order to automatically generate quantum source code in various programming languages, such as QASM, Q#, or Qiskit, among others. The implication of this work should not, however, be understood from the mere point of view of forward engineering. These UML representations may be key aspects during the software modernisation processes employed to migrate classical and quantum software towards hybrid information systems. For example, reverse engineering tools could abstract UML quantum circuits from quantum source code.

This paper proposes the UML extension as part of more extensive long-term research devoted to providing a Quantum UML Profile that will cover other viewpoints and aspects of the analysis and design of hybrid information systems. For example, use case, class, sequence, component and deployment diagrams will be extended with the Quantum UML Profile, and our future research lines will comprise precisely this.

## Acknowledgments

This work is part of the SMOQUIN project (PID2019-104791RB-I00) funded by the Spanish Ministry of Science and Innovation (MICINN) and "QHealth: Quantum Pharmacogenomics Applied to Aging", 2020 CDTI Missions Programme (Center for the Development of Industrial Technology of the Ministry of Science and Innovation of Spain). We would like to thank all the aQuantum members, and particularly Guido Peterssen and Pepe Hevia, for their help and support.